\documentclass[aps,prl,twocolumn,superscriptaddress,showpacs]{revtex4-1}
\usepackage{graphicx}
\usepackage{color}
\usepackage{amsmath}
\newcommand{\commenthe}[1]{}
\renewcommand{\vec}[1]{\mathbf{#1}}
\newcommand{\dosef}{\ensuremath{\mathrm{DOS}_\mathrm{F}}}
\newcommand{\dosunits}{\ensuremath{\,\mathrm{states}/\mathrm{eV}/\text{\AA}^3}}

\begin{document}
\title{How to represent crystal structures for machine learning: towards fast prediction of electronic properties}

\author{K.T. Sch\"utt}\thanks{K.T. Sch\"utt and H. Glawe contributed equally to this work.}\affiliation{Machine Learning Group, Technische Universit\"at Berlin, Marchstr. 23, 10587 Berlin, Germany}

\author{H. Glawe}\thanks{K.T. Sch\"utt and H. Glawe contributed equally to this work.}\affiliation{Max-Planck-Institut f\"ur Mikrostrukturphysik, Weinberg 2, 06120 Halle, Germany}

\author{F. Brockherde}\affiliation{Machine Learning Group, Technische Universit\"at Berlin, Marchstr. 23, 10587 Berlin, Germany}\affiliation{Max-Planck-Institut f\"ur Mikrostrukturphysik, Weinberg 2, 06120 Halle, Germany}

\author{A. Sanna}\affiliation{Max-Planck-Institut f\"ur Mikrostrukturphysik, Weinberg 2, 06120 Halle, Germany}

\author{K.R. M\"uller}\thanks{Corresponding authors; these authors jointly directed the project.}\affiliation{Machine Learning Group, Technische Universit\"at Berlin, Marchstr. 23, 10587 Berlin, Germany}\affiliation{Department of Brain and Cognitive Engineering, Korea University, Anam-dong, Seongbuk-gu, Seoul 136-713, Republic of Korea}

\author{E.K.U. Gross}\thanks{Corresponding authors; these authors jointly directed the project.}\affiliation{Max-Planck-Institut f\"ur Mikrostrukturphysik, Weinberg 2, 06120 Halle, Germany}

\date{\today}

\begin{abstract}
High-throughput density-functional calculations of solids are highly
time consuming. As an alternative, we propose a machine learning approach 
for the fast prediction of solid-state properties. To achieve this, LSDA calculations
are used as training set. We focus
on predicting the value of the density of electronic states at the Fermi
energy.  We find that conventional representations of the input data, such as
the Coulomb matrix, are not suitable for the training of learning machines in the 
case of periodic solids. We propose a novel crystal structure representation
for which learning and competitive prediction accuracies become
possible within an unrestricted class of spd systems of arbitrary unit-cell size.  

\end{abstract}


\maketitle

In recent years \textit{ab-initio} high-throughput computational
methods (HTM) have proven to be a powerful and successful tool to
predict new materials and to optimize desired materials properties. Phase diagrams of 
multicomponent crystals~\cite{Curtarolo2005BiIn,oganov2006review,pickard2011review} and 
alloys~\cite{morgan2005highthroughput} have been successfully predicted. 
High-impact technological applications have been achieved by improving the performance of
Lithium based batteries~\cite{Kang2006lithiumbattery,chen2012lithiumbattery,hautier2013lithiumbattery}, 
by tailoring the non-linear optical response in organic molecules~\cite{keinan2008nonlinearoptical} 
for optical signal processing, by designing desired current-voltage 
characteristics~\cite{Olivares2011organicphotovoltaics} for photovoltaic materials, 
by optimizing the electrode transparency and conductivity~\cite{Peng2013TransparentTopElectrodes} 
for solar cell technology, and by screening metals for the highest amalgamation 
enthalpy~\cite{Jain2010mercurycapture} to efficiently remove Hg pollutants in coal gasification.

However, the computational cost of electronic structure calculations
poses a serious bottleneck for HTM. Thinking of quaternary, quinternary, etc., compounds,
the space of possible materials becomes so large, and the complexity of the unit cells  
so high that, even within efficient Kohn-Sham
density functional theory (KS-DFT), a systematic high-throughput exploration 
grows beyond reach for present-day computing facilities. As a way out, one
would like to have a more direct way to access the physical property of interest 
without actually solving the KS-DFT equations. Machine learning (ML) techniques 
offer an attractive possibility of this type. ML-based calculations are very fast, 
typically requiring only fractions of a second to predict a specific property 
of a given material, after having trained the ML model on a representative 
training set of materials. 

ML methods rely on two main ingredients, the learning algorithm itself
and the representation of the input data. There are many different ways
of representing a given material or compound. While, from the physicist's point of 
view, the information is simply given by the charges and the positions 
of the nuclei, for ML algorithms the specific mathematical form in which
this information is given to the machine, is crucial. Roughly speaking, 
ML algorithms assume a nonlinear map between input data 
(representing the materials or compounds in our case) and the material-specific 
property to be predicted. Whether or not a         
machine can approximate the unknown nonlinear map between input and property 
well and efficiently mainly depends on a good representation \cite{bartok2013representing,von2013representation,braun2008relevant}. 
Recently, ML has contributed accurate models for
predicting molecular properties~\cite{rupp2012fast,montavon2012learning}, transition states~\cite{pardos2010using}, reaction surfaces~\cite{pozoun2012},
potentials~\cite{behler2011atom} and self-consistent solutions for DFT~\cite{snyder2012finding}. All these applications deal 
with finite systems (atoms, molecules, clusters). 
For this type of systems, one particular way of representing the material, 
namely the so-called Coulomb matrix, has been very successful. 

In electronic-structure problems, the single most-important property is 
the value of the density of states (DOS) at the Fermi energy.  
Susceptibilities, transport coefficients, the Seebeck coefficient, 
the critical temperature of superconductors, are all closely 
related to the DOS at the Fermi energy. Therefore,  
we have chosen this quantity to be predicted by ML.

In this work, we shall report a fundamental step forward in the
application of machine learning to predict the DOS at the Fermi energy.
The two main questions this work
aims to address are: (a) How can we describe an infinite
periodic system in a way that supports the learning process well? (b)
How large should the data basis for ML training be, i.e.,  the
\textit{training set} of calculations? Answering these questions will
provide us exactly with the sought-after method of direct and fast prediction
and with the knowledge of whether such prediction is indeed possible
given the finite amount of training data compatible with present
day's computing power.


We employ so-called kernel-based learning
methods~\cite{muller2001introduction,scholkopf2002learning} that are
based on a mapping to a high-dimensional \emph{feature space} such
that an accurate prediction can be achieved with a linear model in
this space.  The so-called \emph{kernel trick} allows to perform this
mapping implicitly using a kernel function, e.g., the Gaussian kernel
$k(\vec{x},\vec{y})= \exp\left(- \| \vec{x} - \vec{y} \|^2 /
\sigma^{2} \right)$  or the Laplacian one $k(\vec{x},\vec{y})=\exp\left(- \| \vec{x} - \vec{y} \| /
\sigma \right) $. Kernels can be viewed as a similarity measure
between data, in our case they should measure proximity between
materials for a certain property.  The property to be predicted is
computed as a linear combination of kernel functions of the
material of interest and the training materials.  Therefore, constructing a
structure representation in which crystals have small distance when
their properties are similar is beneficial for the learning process (see below for details). 

In order to predict the DOS, we employ \emph{kernel ridge regression (KRR)}, which is a kernelized
variant of least-squares regression with $\ell_2$-regularization.
Additionally, the predictive variance can be estimated, which can serve as a measure of how well a material of interest is represented in the training set.
We use nested cross-validation for the model selection process~\cite{trevor2001elements,hansen2013assessment}, i.e., the parameter selection and performance
evaluation are performed on separate held-out subsets of the data that are
independent from the set of training materials.  This ensures to find
optimal parameters for the kernel and the model regularization in terms of
generalization while avoiding overfitting. 

In the solid state community crystals are conventionally described by
the combination of the \textit{Bravais Matrix}, containing the
primitive translation vectors, and the \textit{basis}, setting the
position and type of the atoms in the unit cell.  This type of
description is not unique and thus not a suitable representation for the learning process
since it depends on an arbitrary choice of the coordinate system in
which the Bravais matrix is given. Namely, there exists an infinite
number of equivalent representations that would be perceived as
distinct crystals by the machine.  In principle, recognizing
equivalent representations could also be tackled by machine learning
directly as done for molecules in Ref.~\cite{montavon2012learning, montavonNJP13, montavonSPM13}.
However, a significant computational cost in terms of size of the
training set had to be paid.  
Due to the aforementioned larger ambiguity in the case of crystals,
an even higher cost is expected.

For the case of molecules the \emph{Coulomb matrix} has proven to be a
well-performing
representation~\cite{rupp2012fast,montavon2012learning}.  This is
given by
\[
  C^{\mathrm{mol}}_{ij} = \begin{cases}
      0.5Z_i^{2.4} & \text{for } i = j\\
      \frac{Z_i Z_j}{\|\vec{r}_i - \vec{r}_j\|} & \text{for } i \neq j
  \end{cases} 
\]
with nuclear charges $Z_i$ and positions $\vec{r}_i$ of the
atoms. This description is invariant under rotation and translation, but
unfortunately it cannot be applied directly to infinite periodic
crystals.

A simple extension to crystals is to combine a Coulomb matrix of
one single unit cell with the Bravais matrix.  
However this representation suffers of the \textit{degeneracy problem} 
mentioned above.
The Coulomb matrix representation assumes a similarity relation between
atoms with close nuclear charges.  However, this is most often not
the case for the chemical properties. 

\begin{figure}[htbp]
\includegraphics[width=0.4\columnwidth]{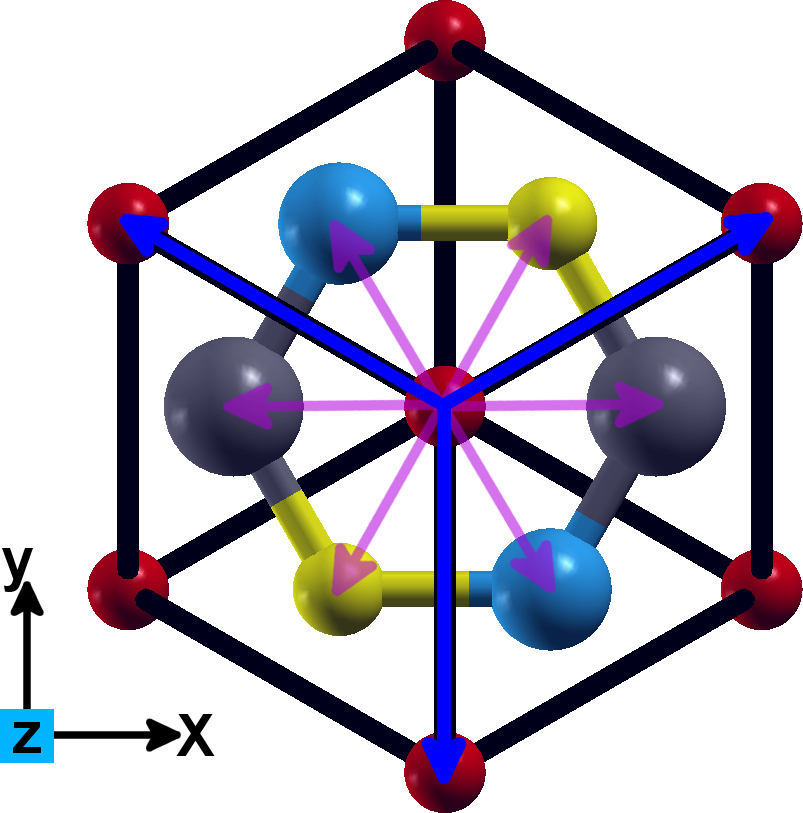}
\includegraphics[width=0.4\columnwidth]{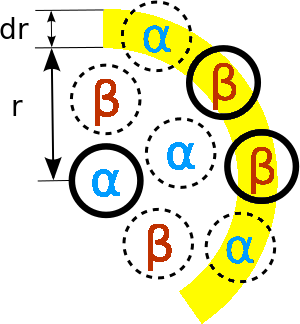}
\caption{Alternative crystal representations. Left: a crystal unit cell with indicated the Bravais vectors (blue) and base (pink). Right:
Illustration of one shell of the discrete partial radial distribution function $g_{\alpha\beta}(r)$ with width $dr$.\label{fig:prdfillu}}
\end{figure}

In order to include more physical knowledge about crystals, we propose a novel crystal representation inspired by radial distribution functions as used in the physics of x-ray powder diffraction~\cite{billinge1998local}
and text mining from computer science~\cite{forman2003extensive,joachims1998text}.
The \textit{partial radial distribution function (PRDF) representation} considers the distribution of pair-wise distances $d_{\alpha \beta}$
between two atom types $\alpha$ and $\beta$, respectively.  This can
be seen as the density of atoms of type $\beta$ in a shell of radius
$r$ and width $dr$ centered around an atom of type $\alpha$ (see
Fig.~\ref{fig:prdfillu}).  Averaged over all atoms of a type, the
discrete \textit{(PRDF) representation} is given by
\[
g_{\alpha\beta}(r) = \frac{1}{N_\alpha V_{r}} \sum_{i=1}^{N_\alpha} \sum_{j=1}^{N_\beta}  \theta(d_{\alpha_i \beta_j} - r) \theta(r + dr - d_{\alpha_i \beta_j}),
\]
where $N_\alpha$ and $N_\beta$ are the numbers of atoms of type
$\alpha$ and $\beta$, respectively, while $V_{r}$ is the volume of the
shell.  We only need to consider the atoms in one primitive cell as
shell centers for calculation.  The distribution is globally valid due
to the periodicity of the crystal and the normalization with respect
to the considered crystal volume.  In this work, the type criterion
for 'counting' an atom is its nuclear charge, however, other more
general criteria could in principle also be used, such as the number of valence
electrons or the electron configuration. 

As input for the learning algorithm, we employ the feature matrix X
with entries $x_{\alpha\beta,n} = g_{\alpha\beta}(r_n)$, i.e., the
PRDF representation of all possible pairs of elements as well as
shells up to an empirically chosen cut-off radius.  The distance of two
crystals is then defined as the distance induced by the Frobenius norm
between those matrices and may be plugged into one of the previously
described kernels. In this manner, we have defined a novel
global descriptor as well as a similarity measure for crystals which is
invariant under translation, rotation and the choice of the unit cell.

The \dosef\ we use to train and validate the learning are
computed~\footnote{All calculations are performed within KS spin
  density functional thory~\cite{KS,DFT}, with LSDA xc~\cite{LSDA}. Core
  states are accounted in the pseudo-potential approximation as
  implemented in the ESPRESSO package~\cite{QE-2009}
  $k$-points are sampled with a Monkhorst-Pack
  grid~\cite{MonkhorstPack} with a density of about
  $1510\mathrm{\AA}^3$. Magnetic ordering is assumed to be ferro-magnetic.} on
crystals from the inorganic crystal structure database
(ICSD)~\footnote{ICSD, Inorganic Crystal Structure Database,
  Fachinformationszentrum Karlsruhe: Karlsruhe, Germany, (2011).11}
with the experimental lattice parameters reported therein. The chosen
subset contains only non-duplicated materials with a maximum of 6
atoms per primitive cell.  We subdivide the set into $sp$ (1716
crystals) and $spd$ (5548 crystals).



\begin{table}[htbp]
  \caption{Mean absolute errors and standard errors of DOS predictions in $10^{-2}\dosunits{}$\label{tab:mae}}
\begin{ruledtabular}
\begin{tabular}{llrr}
\textbf{Predictor}  & \textbf{Features} & \textbf{sp systems} & \textbf{spd systems} \\ \hline
Mean predictor     & -- & $1.50 \pm 0.02$            &  $1.82 \pm 0.03$ \\ \hline
KRR (linear)	 & B+CM & $1.45 \pm 0.04$ 			& $1.68 \pm 0.01$ \\
KRR (gauss.)     & B+CM & $1.19 \pm 0.03$           &   $1.62 \pm 0.01$   \\ 
KRR (lapl.)    & B+CM & $1.20 \pm 0.04$      &   $1.63 \pm 0.02$   \\ \hline
KRR (linear)	 & PRDF & $0.87 \pm 0.02$ 			& $1.68 \pm 0.03$ \\
KRR (gauss.)     & PRDF & $0.74 \pm 0.03$       &   $0.95 \pm 0.02$   \\ 
KRR (lapl.)    & PRDF & $\textbf{0.68} \pm 0.03$   &   $\textbf{0.86} \pm 0.01$   \\
\end{tabular}
\end{ruledtabular}
\end{table}

\begin{figure}[htbp]
\includegraphics[width=0.8\columnwidth]{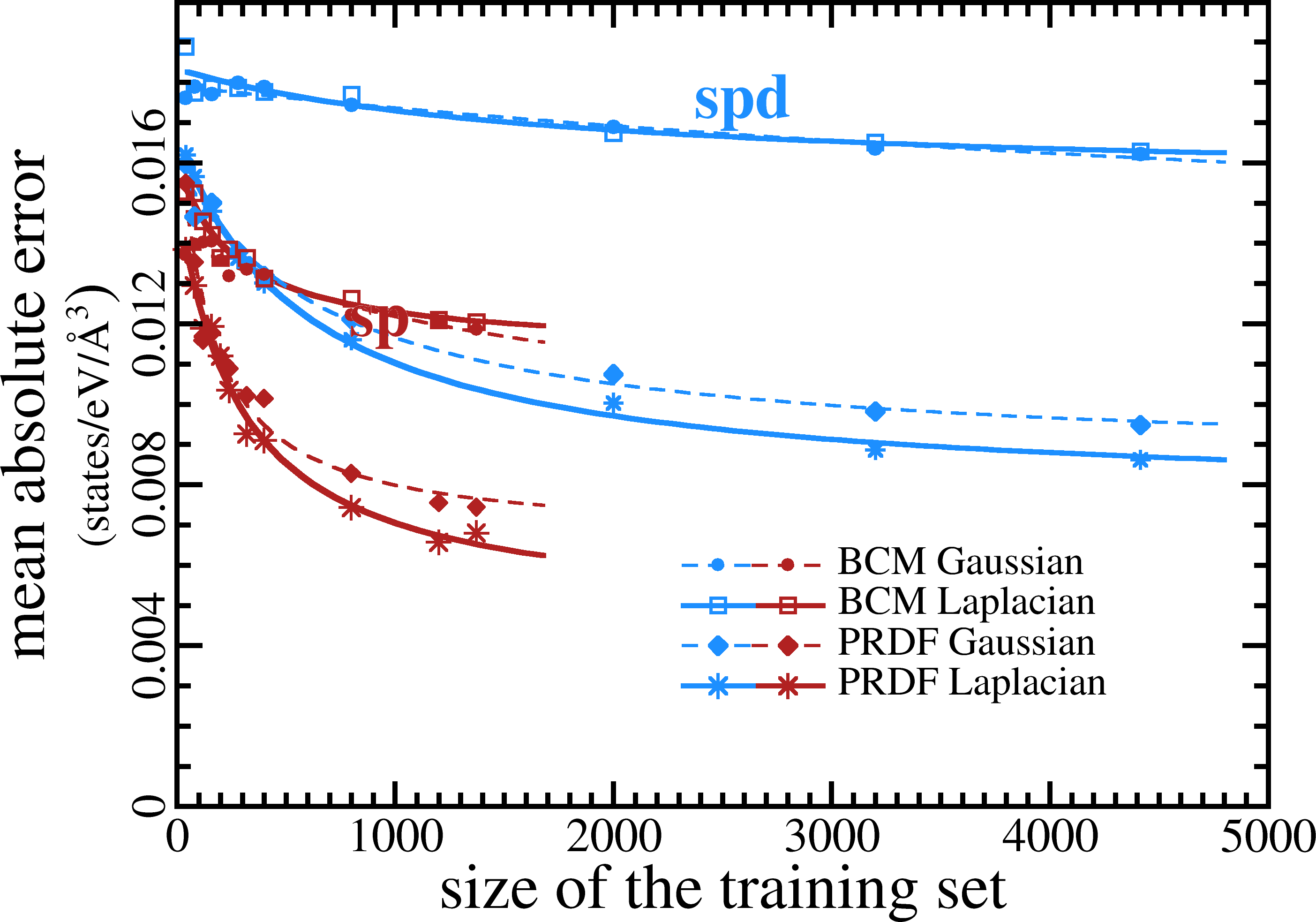}
\caption{Learning process as a function of the number of materials used for training for all three feature representations (conventional CCM and PRDF), and for the two datasets.\label{fig:learningcurve}}
\end{figure}

For the \dosef\ prediction, we first consider the sp and spd 
material sets separately.  The mean absolute errors of the predictions of all
presented crystal representations are collected in Table
\ref{tab:mae}.  Furthermore, we list the mean predictor that always
predicts the average \dosef\ value of the training set as a simple
baseline. Both representations yield models that are significantly better than the mean predictor.
Fig.~\ref{fig:learningcurve} illustrates how the error
decreases steadily with increasing number of materials used for
training. 
However, the PRDF features consistently outperform the B+CM description. The further analysis will therefore focus on PRDF with the slightly better performing Laplacian kernel.

The higher complexity of the spd systems can clearly be observed in
the learning curves, which show how much better the prediction problem
can be solved as a function of the available data. The mean error is
much lower in sp materials. Furthermore, the learning curves are
steeper, i.e., increasing the training set size within the restricted
materials class improves the prediction accuracy rapidly.  One origin
of this higher complexity lies in the growing dimensionality of the
input space: given $N_\mathrm{el}$ possible chemical elements in all
material compositions, $\dim(\text{X})\propto N_\mathrm{el}^2$.
Furthermore, by including materials with d electrons, the physics
becomes more rich.  Due to both reasons, much more training data is
required to achieve an improvement comparable to that of sp systems.

\begin{figure}[tbp]
\begin{minipage}{0.8\columnwidth}
\includegraphics[width=\columnwidth]{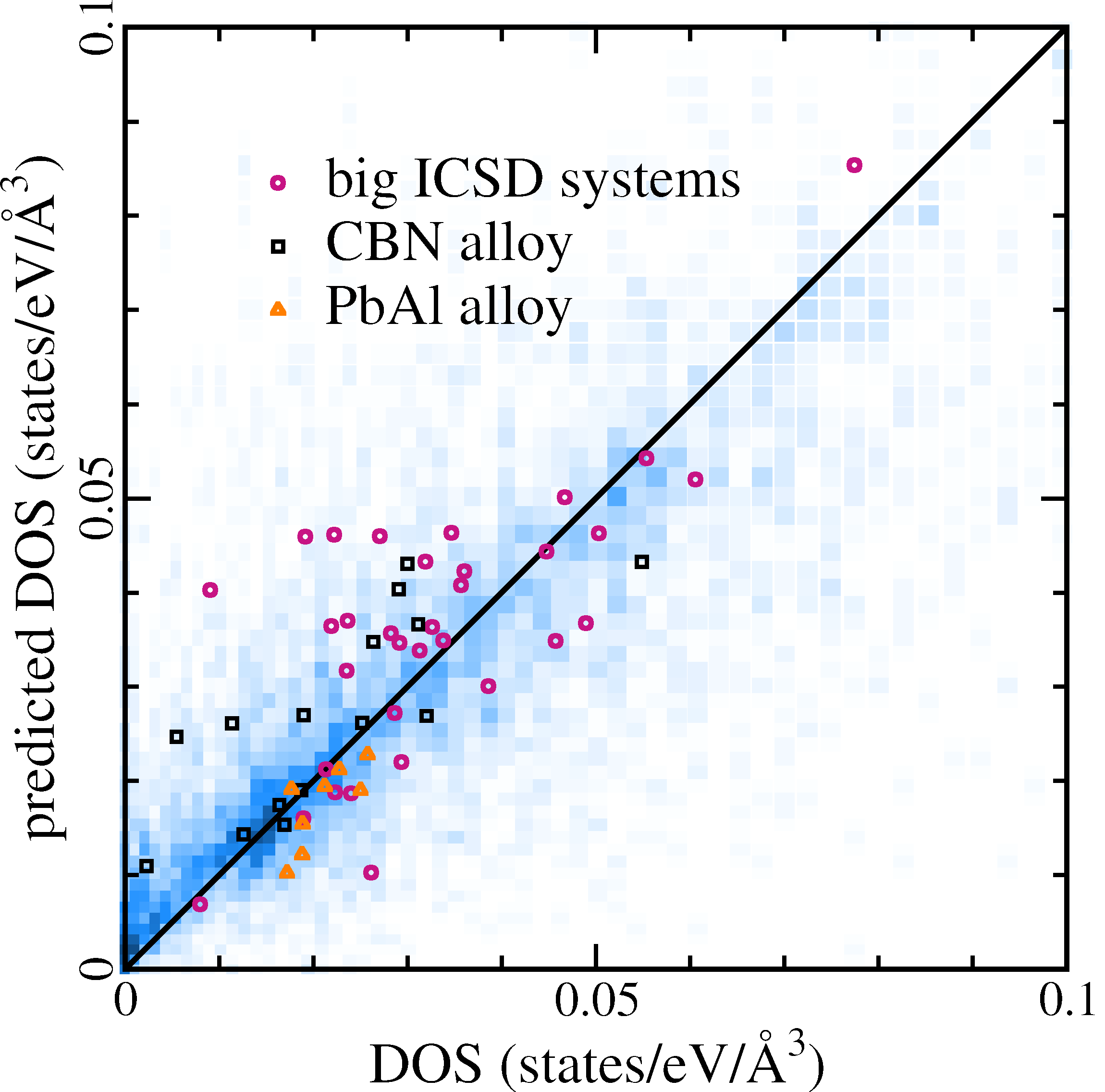}
\end{minipage}
\caption{Comparison between predicted and calculated \dosef\ for spd systems. The background density distribution refers to the cross validation systems. Dots are additional systems (see legend) of far larger size than those used for training. \label{fig:IV}}
\end{figure}

The prediction of \dosef\ for spd systems is shown in Fig.~\ref{fig:IV}, as a density plot of computed versus predicted values.  
It is evident that the density is accumulated along the diagonal of the plot,
demonstrating that the machine is giving meaningful predictions. 
While the average error is smaller than 6\% of the DOS value range. Thus this result represents a proof of principle that 
a complex output of the Kohn-Sham equations can be predicted directly by means of machine learning -- albeit the considerable variance of errors. 

From Fig.~\ref{fig:learningcurve}, it is clear that, in order to increase the prediction accuracy, the size of the training set should be extended, possibly at the limits of present computing facilities. Instead of a brute-force approach, the problem may become less costly by using an active learning scheme, e.g., by extending the set where the
\textit{predicted variance} is higher. We still expect that in order to obtain highly accurate results the computation costs will be large.
Can the proposed approach still be useful at the present accuracy level?

To answer this question we first point out that ML is at least 2 or 3 orders of magnitude faster than any direct computational approach, this means that a fast ML scan may always be of use as a preliminary step before consuming precious resources on detailed calculations.
Second, a remarkable feature of the PRDF representation is that it is not fixed to a certain number of atoms in the unit cell of the training materials. This means that once the machine has been trained, it can be used to predict the properties of any other system. This is virtually independent from its size as long as it is well represented by the training set. 

As a proof for this ability, we consider additional test systems, divided into 3 set: The first set (pink circles in Fig.~\ref{fig:IV}) contains only systems taken from the ICSD database, chosen among those with between 30 and 80 atoms per unit cell that are well represented by the training set. Therefore, only ICSD materials with a relatively low predictive variance were chosen for calculation. The second set (orange triangles in Fig.~\ref{fig:IV}) contains a purely metallic alloy (within the unit cell) of lead and aluminum. All the systems in this set are crystals with 125 atoms per unit cell, differing by the Al/Pb concentration. The third set (black squares in Fig.~\ref{fig:IV}) is a solid solution of three atomic types, in a diamond lattice: Carbon, Boron and  Nitrogen, at different composition and a total of 45 atoms per periodic unit cell.

Unlike the training systems, each of these involve a large computational cost, and would not be feasible without access to a computation facility.
While the PbAl alloys are quite well predicted, some of the \dosef\ for the large ICSD systems (mostly oxides) are overestimated, as well as some of the \dosef\ of the CBN solid solution. Again, a clear diagonal accumulation is achieved. Nevertheless for these large systems that the learning machine has never been trained on, the average quality of the prediction of large systems is comparable to that of the much smaller, cross-validated systems.

In summary, we have investigated a machine learning approach for fast solid-state
predictions. A set of LSDA calculations has been used to 
train a \dosef\ predictor.
We expect that our method can be extended to directly predict other complex
materials properties as well.
It certainly can be combined with other, more accurate, electronic structure techniques such as $GW$.
The accuracy of predictions depends strongly on how crystals are represented.
We found that Coulomb matrices, while being successful for predicting
properties in molecules~\cite{montavonNJP13,montavonSPM13}, are not suitable to
describe crystal structures well enough. Instead, we have proposed a
representation inspired by partial radial distribution
functions which is invariant with respect to translation, rotation and the choice of the unit cell.
Our results clearly demonstrate that a fast prediction of electronic properties in solids with ML algorithms
is indeed possible.
Although presently the accuracy leaves room for improvement, we consider the predictions useful for a first screening of a huge number of materials for properties within a desired value range. In a second step, high-accuracy electronic structure calculations are then performed on the promising candidates only.
What makes the approach extremely appealing is that the PRDF representation allows to learn on small systems with low computational cost and then extrapolate to crystals with arbitrary numbers of atoms per unit cell for which conventional DFT calculations would be prohibitive.

\begin{acknowledgments}
F. Brockherde and K.-R. M\"uller gratefully acknowledge helpful discussions with Matthias Scheffler, Claudia Draxl, Sergey Levchenko and Luca Ghiringhelli, who pointed out to us that for electron densities and band gaps the local topology and connectivity of the atoms is an appropriate descriptor and not the Coulomb matrix.
Furthermore we acknowledge valuable comments of Alexandre Tkatchenko, Katja Hansen and Anatole von Lilienfeld. KRM, KS and FB thank the Einstein Foundation for generously funding the ETERNAL project. 
\end{acknowledgments}

\bibliography{literature}
\end{document}